\documentclass[amsmath, amssymb,12pt]{article}
\title{\vspace*{-1cm}{\normalsize\rightline{CU-TP-1128}}
{\Large  \bf  A Possible Modified ``bottom-up'' Thermalization in Heavy Ion Collisions}\footnote{This work is supported in part by the US Department of Energy}}

\author{}
\date{}
\begin{document}
\maketitle

\vspace*{-2cm}
\begin{center}

\renewcommand{\thefootnote}{\alph{footnote}}

{\large A.~H.~Mueller\,\footnote{arb@phys.columbia.edu (A.H. Mueller)},
 A.~I.~Shoshi\,\footnote{shoshi@phys.columbia.edu} and 
S.~M.~H.~Wong\footnote{s\_wong@phys.columbia.edu}}

\vspace*{0.5cm}

{\it Department of Physics, Columbia University, \\
New York, New York  10027, USA}

\end{center}


\begin{abstract}

\indent We present a possible scaling solution to pre-equilibrium evolution which interpolates between the instability present in the dense gluon system produced immediately after a heavy ion collision and the final equilibration which occurs later.  Our solution depends on a single parameter $\delta$.  Depending on the value of $\delta$, our proposed solution matches onto the bottom-up picture either at an intermediate stage or toward the end of the evolution given by bottom-up.  We discuss in detail the reasons why we believe our solution is self-consistent, and we also point out why it is difficult to actually prove consistency.
\end{abstract}

\bigskip
\section{Introduction}

The bottom-up picture[1] of thermalization in heavy ion reactions was an attempt to describe the various stages that the dense system of gluons produced in a heavy ion reaction go through on the way to equilibration.  The picture, which 
relied on the observation that inelastic processes were important for kinetic equilibration\cite{wong}, was relatively simple with three distinct stages. An early stage dominated by hard gluons having momenta on the order of $Q_s,$ the saturation momentum of the colliding nuclei; an intermediate stage where hard gluons dominate the total number of gluons but softer gluons dominate the Debye mass calculation; a final stage where the soft gluons are thermalized and receive energy, by QCD branching, from the hard gluons until all the energy is removed from the hard gluons and the whole system is equilibrated.

Arnold, Lenaghan and Moore[3] pointed out a severe problem with the bottom-up picture.  Rather than having a screening mass at early times the system of gluons produced has a very asymmetric momentum distribution which leads to a mass characterizing an instability[3-7].  Thus, the early stage of the bottom-up picture is not tenable.  The instability clearly leads to an increase in the transfer of energy from hard to soft gluons, and so one could hope that it produces a rapid thermalization.  However, Arnold and Lenaghan[8] showed that complete equilibration cannot occur before $Q_s\tau = \alpha^{-7/3}$ a time not much shorter than that which occurs in the bottom-up picture.  Thus the instability cannot lead directly to equilibration since that would give an equilibration time parametrically on the order of $Q_s\tau\sim\;1.$  Thus another mechanism might be necessary to fill this gap.  There are no lack of proposals in the literature for this ranging from a strongly coupled quark-gluon plasma[9,10], a mechanism using the Hawking-Unruh effect[11] and collinear enhancement within the bottom-up picture[12].  No matter what the mechanism really is as long as it is perturbative in nature\footnote{Clearly not all the above mechanisms are perturbative.}, it would be natural to look for a scaling solution which begins when the instability runs its course and ends when equilibration occurs.  And, of course, this scaling solution cannot be the bottom-up solution. 

In this paper we search for solutions to the Boltzmann equation where the collision term is given in terms of two to two and two to three gluonic processes.  We allow both the hard gluons, the remnants of the gluons produced in the initial ion-ion collision, and the soft gluons, produced later on in the evolution of the system, to be part of the collision term.  As in Ref.1 we treat the collision term in a dimensional way without trying to keep track of constant and logarithmic prefactors to the dominant behaviors.

We have found a one-parameter family of scaling solutions parametrized by a positive number $\delta.$  We believe that these are the only scaling solutions which exist.  They are exhibited in Eq.(1) for $N_s$ the number density of soft gluons, $k_s$ the soft gluon momentum, $f_s$ the soft gluon occupation number and $m_D$ the Debye screening mass.  Unfortunately, we cannot prove that $m_D$ is actually a screening mass rather than an instability mass.  We show in Sec. 2 that our proposed solution only has internal consistency if $m_D$ is interpreted as a screening mass.  As we believe the system should have a scaling solution connecting the instability to equilibration, and we believe the solutions in (1) are the only possible  scaling solutions, we are hopeful, but not absolutely sure, that the $m_D$ in (1) is a genuine screening mass.  At the end of Sec. 2 we discuss the competing mechanisms which occur in trying to decide whether $m_D$ is screening or whether it corresponds to an instability.

In Sec. 3 we show that our proposed solution matches onto the intermediate stage of the bottom-up picture for $0 < \delta < 1/3,$ with $\delta = 0$ being the bottom-up scenario and $\delta = 1/3$ having the matching occur just as the soft particles are beginning to equilibrate.  Solutions with $\delta > 1/3$ create equilibration of soft particles earlier than the corresponding time in bottom-up, and in this case the matching with bottom-up only happens as the whole system is reaching equilibration. Although the thermalization we find here is parametrically the same as in bottom-up the system may well become ``effectively thermalized'' at an earlier time as suggested in Ref.[6].

As in bottom-up equilibration occurs when $Q_s\tau\sim \alpha^{- 13/5}$, independent of $\delta.$  The fact that we have a family of solutions rather than a single solution reflects the weakness of our approach.  An approach utilizing more dynamics, such as an actual solution to the Boltzmann equation, should find a unique solution matching onto the instability at early times.

Finally, in Sec. 4 we give a simple proposal of when and how the instability may run its course.

\section {The scaling solution}

Our procedure here is just to present a solution for the evolution of the pre-equilibrium gluon system, after the instability has run its course, and then to check that this solution appears to satisfy all the requirements that such a solution should satisfy.  The part of the following discussion which is somewhat unsatisfactory is our inability to prove that what we call the Debye mass, $m_D,$ actually obeys $m_D^2>0.$  At the end of our discussion in this section we comment as to why this is not so easy to show.

Our proposed solution is given by

\begin{displaymath}
N_s\sim{Q_s^3\over \alpha(Q_s\tau)^{4/3-\delta}},\quad k_s\sim {Q_s\over (Q_s\tau)^{1/3-2\delta/5}},
\end{displaymath}
\begin{equation}
 \alpha f_s\sim {1\over (Q_s\tau)^{1/3+\delta/5}},\quad  m_D\sim {Q_s\over(Q_s\tau)^{1/2-3\delta/10}}
\end{equation}

\noindent with $\delta$ a positive real number and where, as in Ref.[1], the symbol $\sim$ means that we are unable to evaluate the constant factors  (perhaps logarithmic factors) which accompany the dimensional factors, the powers of $\alpha$ and the powers of $Q_s\tau$ which we have evaluated.  Again, as in Ref.[1], we separate hard gluons having momentum on the order of $Q_s$ from the ``soft'' particles produced around time $\tau$ which have momentum  $k_s.$  In addition, there are of course gluons which have been produced in the interval $1/Q_s \ll \tau_0\ll \tau$ and these gluons, produced at $\tau_0,$ have

\begin{displaymath}
N_s(\tau, \tau_0) \sim {Q_s^3\over \alpha(Q_s\tau)(Q_s\tau_0)^{1/3-\delta}},\quad  k_s(\tau_0)\sim {Q_s\over (Q_s\tau_0)^{1/3-2\delta/5}},
\end{displaymath}
\begin{equation}
\alpha f_s(\tau, \tau_0) \sim {(Q_s\tau_0)^{1/3+\delta/5}\over (Q_s\tau)^{2/3 + 2\delta/5}}
\end{equation}

\noindent where, for example, $N_s(\tau,\tau_0)$ stands for particles produced around time $\tau_0$ and measured at time $\tau > \tau_0.$  In evaluating $\alpha f_s(\tau,\tau_0)$ we have used

\begin{equation}
f_s(\tau,\tau_0) \sim {N_s(\tau,\tau_0)\over k_s^2(\tau_0)k_s(\tau)}
\end{equation}

\noindent because the soft particles produced at $\tau_0,$ just as the hard particles, have a longitudinal momentum determined by multiple scattering at time $\tau.$  Since $N_s(\tau,\tau_0) f_s(\tau,\tau_0) < N_sf_s$ the particles produced at $\tau_0$ are not so important for scattering which occurs at time $\tau,$ and so for the present we neglect them.  At the end of this section we will come back to the role they play, and we shall see that the separation of $N_s$ from $N_s(\tau,\tau_0)$ is subtle.

The solution (1) has the following properties:

\begin{equation}
m_D^2\sim {\alpha N_s\over k_s}
\end{equation}
\begin{equation}
N_s\sim \tau{\alpha^3\over m_D^2}(N_sf_s)^2
\end{equation}
\begin{equation}
k_s^2\sim m_D^2{\tau\over \tau_s^{col}}\quad  {\rm with}\quad  {1\over \tau_s^{col}} \sim {\alpha^2\over m_D^2} N_sf_s
\end{equation}
\begin{equation}
{1\over \tau} \sim {\alpha^2\over k_s^2} N_sf_s.
\end{equation}

In (5)-(7) we have included the usual final state occupation factors in the approximation $1+ f_s\simeq f_s.$  These occupation factors are essentially identical to those appearing in the early stages of the bottom-up picture[1]

\noindent Eq.(4) says that the soft particles dominate the Debye mass calculation and that they do so in a self-consistent manner.  Eq.(5) says that the soft particles are produced from other soft particles (produced a little earlier) by the Bethe-Heitler process.  Eq.(6) says that once particles are produced at a momentum $m_D,$ by the Bethe-Heitler process, they then multiply scatter with soft particles produced a little earlier to reach the momentum $k_s$ given in (1).  Finally, Eq.(7) says that a soft particle scatters with momentum transfer $k_s,$ with other soft particles once in the time period $\tau.$  Thus, these particles are at the edge of equilibration.

Now, since we have not been able to definitively determine that the $m_D$ in (1) is a screening mass rather than instability mass why have we assumed it to be a screening mass?  The answer is straightforward; it is not internally consistent to  suppose the $m_D$ of Eq.(1) to be an instability mass.  To demonstrate this we need to calculate $\tau_m,$ the time over which particles produced at a scale $m_D$ stay at that scale before having their momentum increased (by say a factor of 2), and $f_m$ the occupation number of particles at the scale  $m.$  Clearly

\begin{equation}
{1\over \tau_m} \sim {\alpha^2\over m_D^2} N_sf_s = m_D{1\over (Q_s\tau)^{1/6 +\delta/10}}
\end{equation}

\noindent and

\begin{equation}
\alpha f_m  \sim \alpha N_s {\tau_m\over \tau}{1\over m_D^3} = {1\over (Q_s\tau)^{1/6 +\delta/10}}.
\end{equation}
	
\noindent The fact that $\tau_m\gg 1/m_D$ while $\alpha f_m\ll 1$ means that the particles at $m_D$ do not come from an instability, because the instability would quickly raise $\alpha f_m $ to a value at least on the order of 1.

Finally, what is the difficulty in deciding whether the particles described by Eq.(1) have a sufficiently symmetrical distribution so that screening dominates instability?  The heart of the problem can be seen using Eq.(2) to calculate the ``Debye mass'' at $\tau$ due to particles produced at $\tau_0.$  One finds

\begin{equation}
m_D^2(\tau,\tau_0) \sim {\alpha N_s(\tau,\tau_0)\over k_s(\tau_0)} \sim {
Q_s^2(Q_s\tau_0)^{3\delta/5}\over (Q_s\tau)}.
\end{equation}

\noindent Of course if $\tau_0/\tau \ll 1, m_D^2(\tau,\tau_0)$ will be an instability mass, because the particles produced at $\tau_0$ will have a very asymmetric momentum distribution at  $\tau$ [3].  But then if $\tau_0/\tau \ll 1, m_D^2(\tau,\tau_0)\ll m_D^2(\tau)$ so that these gluons are not important.  The gluons produced very close to the time $\tau$ have equilibrating interactions among themselves and their distribution may well be quite symmetric.  What is difficult to decide is the role of gluons with, for example, $\tau_0=1/2 \tau.$  These gluons have  begun to have an asymmetric distribution and their contribution to $m_D^2$ is  not much less, depending on $\delta,$ than the gluons produced at $\tau.$  Thus, there are likely various contributions to $m_D^2$ some screening and some favoring instability.  Larger values of $\delta$ favor screening because the gluons produced later are more emphasized, however, we have not been able to find a technical argument which decides whether screening or instability dominates for a particular $\delta.$

\section{Matching onto the bottom-up picture}

As observed in Ref.[3] the early time $(1< Q_s\tau < \alpha^{-3/2})$ part of the bottom-up picture is not correct because what was assumed to be a Debye mass was in fact a scale characterizing an instability.  The late time $(\alpha^{-5/2}< Q_s\tau)$ part of the bottom-up picture should be consistent because the Debye mass was there determined from thermalized soft gluons. It is not immediately obvious whether at intermediate times $(\alpha^{-3/2} < Q_s  \tau < \alpha^{-5/2})$ the bottom-up scenario is self-consistent.  Here the interaction of the soft particles among themselves is of the same strength as the soft particles with the hard particles.  In each case the time scale for an elastic scattering with momentum transfer on the order of  $k_s^{bottom-up} \sim Q_s{\sqrt{\alpha}}$ is the time $\tau$ itself.  We do not know how to answer the question of the stability of the intermediate time region of the bottom-up scenario although this issue is crucial for determining exactly how our present solution matches onto the bottom-up picture.

Our present solution is in fact a family of solutions, depending on the parameter $\delta.$  We now observe that the solution given in Eq.(1) agrees with the corresponding intermediate time quantities of the bottom-up scenario at a single time given by

\begin{equation}
Q_s\bar{\tau}=\left({1\over \alpha}\right)^{15\over 10-12\delta}
\end{equation}

\noindent when $0\le \delta \le 1/3.$  For reference we recall that when $\alpha^{-3/2}< Q_s\tau < \alpha^{-5/2}$ the bottom-up quantities are

\begin{equation}
N_s\sim {\alpha^{1/4}Q_s^3\over (Q_s\tau)^{1/2}},\quad  k_s\sim{\sqrt{\alpha}}Q_s, \quad \alpha f_s\sim {1\over \alpha^{1/4}(Q_s\tau)^{1/2}},\quad  m_D\sim {\alpha^{3/8}Q_s\over (Q_s\tau)^{1/4}}.
\end{equation}

\noindent Thus when $\tau=\bar{\tau}$ there should be a transition from our present solution, Eq.(1), to the bottom-up solution, Eq.(12).  For $\tau > \bar{\tau}$ the bottom-up solution should be the correct one.  We note that in case $\delta = 1/3$  the transition is to the beginning of the final stage of bottom-up, starting at $Q_s\tau \sim\alpha^{-5/2}.$  Of course this transition to bottom-up can  occur, for $\delta < 1/3,$ only if the intermediate time part of bottom-up is internally consistent.  If this is not the case then $\delta$ must be greater or equal to $1/3.$

If $\delta > 1/3$ the solution (1) changes character at a time $\tau_1$ given by

\begin{equation}
Q_s\tau_1\sim (1/\alpha)^{15\over 5+3\delta}
\end{equation}

\noindent at which time $f_s\sim 1.$  In this case our solution goes into an evolution much like the final phase of bottom-up where the soft gluons are thermalized and the harder gluons feed energy into the soft thermalized system causing the temperature to rise with time until the whole system is thermalized.  The difference now is that $N_s(\tau,\tau_0)$ is larger than the number of hard particles so that as time grows larger and larger the gluons from $N_s(\tau,\tau_0)$ at smaller and smaller  $\tau_0$ disappear into the thermal bath, through branching as in bottom-up.  The equations governing this evolution are

\begin{equation}
{d\epsilon\over d\tau}\sim T^3{dT\over d\tau}\sim {N_s(\tau,\tau_0)\over \tau}\cdot k_s(\tau_0)
\end{equation}

\noindent where

\begin{equation}
k_s(\tau_0)\sim \alpha^4T^3\tau^2
\end{equation}

\noindent Eq.(15) is the same as in bottom-up where now $k_s(\tau_0)$ plays the role of $k_{br}$ while Eq.(14) is the same as (14) and (15) of the first paper in Ref.[1] when one replaces $N_h$ of that reference by $N_s(\tau,\tau_0).$  Using (15) to give the dependence of $\tau_0$ on $T(\tau)$ and $\tau$ one finds, using (2),

\begin{equation}
T\sim Q_s \alpha^{35-78\delta\over 39\delta-10}\ (Q_s\tau)^{15-36\delta\over 39\delta-10}.
\end{equation}

\noindent It is straightforward to verify that when $\delta=1/3,  T \sim \alpha^3Q_s^2\tau$ as in Ref.[1] and also that $T^4\sim {Q_s^4\over \alpha(Q_s\tau)},$ the endpoint of the non-equilibrium evolution, when $Q_s\tau = \alpha^{-13/5}$ with $T=Q_s\alpha^{2/5}$ independent of the values of $\delta.$  Thus when $\delta > 1/3$ the matching onto bottom-up occurs only at the final time, $Q_s\tau\sim\alpha^{-13/5}.$

How big could $\delta$ possibly be?  Certainly $\delta < 10/21$ because when $\delta={10\over 21}, N_sk_s$ is equal to the energy density of the hard particles and this is an absolute limit as to the amount of energy carried by the soft gluons.

\section{What happens to the instability?}

So far we have said little about the instability except to comment that it should turn into something else in a very short time.  Here we try to be a little more quantitative starting from the Abelianization picture of Arnold and Lenaghan[8] who suggest, and give evidence, that the instability should grow in one color direction.  Let $f_1$ be the gluon occupation in that Abelian direction at a momentum scale $m,$ the instability mass.  Of course when there are many gluons in a single color direction they may scatter elastically into other color directions through a lowest order collision term.  Call the occupation number in these other color directions, but still at a momentum scale $m, f_2.$  Then

\begin{equation}
{df_2\over d\tau} \sim (m^3f_1) {\alpha^2\over m^2}[f_1(1+f_2)].
\end{equation}

\noindent The first term on the right-hand side of (17) is just the number of gluons in the Abelian direction, the second factor is the cross section for scattering into the other color directions, and the third factor gives the remaining boson factors.  Write (17) as

\begin{equation}
{1\over 1  + f_2} \ {d f_2\over d\tau} \sim m (\alpha f_1)^2.
\end{equation}

\noindent Now we suppose that $f_1(\tau) \sim e^{m\tau}f_1(0).$  This gives

\begin{equation}
f_2(\tau) \sim [\alpha f_1(\tau)]^2
\end{equation}

\noindent so long as $\alpha f_1$ is small.  The lowest order collision term which leads to (19) is only systematic so long as $\alpha f_1 < 1.$  Thus, although (19) allows one to follow the rise of $f_2$ to become a quantity of order one it is not possible to reliably follow $f_2$ into the region where it becomes much greater than 1, although the elastic collision terms themselves clearly cause $f_2$ to become as large as $f_1.$  Our proposed solution supposes that the instability is damped by a growth of soft gluons in all color directions and that there is then a transition to a solution of the form which we have presented. In this regard the recent numerical calculations by Dumitru and Nara[13] are interesting in that they show that the Abelian instability seems to run its course before there is a significant change in the hard particle distribution.  It may well be that what is being seen is a damping of the instability due to soft gluon production of the type indicated in (17).

\vskip 10pt
\noindent{\bf Acknowledgements}
\bigskip

\noindent This paper came out of an earlier collaboration with Rolf Baier, Dietrich B\"odeker and Dominique Schiff whose influence on this work we gratefully acknowledge.  A. Shoshi acknowledges financial support by the Deutsche Forschungsgemeinschaft under contract Sh 92/1-1.

\end{document}